# Spin Textures in Synthetic Antiferromagnets: Challenges, Opportunities, and Future Directions

Kang Wang*, Vineetha Bheemarasetty, Gang Xiao*

*Department of Physics, Brown University, Providence, Rhode Island 02912, USA*

## Abstract

Spin textures such as magnetic domain walls and skyrmions have the potential to revolutionize electronic devices by encoding information bits. Although recent advancements in ferromagnetic films have led to promising device prototypes, their widespread implementation has been hindered by the material-related drawbacks. Antiferromagnetic spin textures, however, offer a solution to many of these limitations, paving the way for faster, smaller, more energy-efficient, and more robust electronics. The functionality of synthetic antiferromagnets, comprised of two or more magnetic layers separated by spacers, may be easily manipulated by making use of different materials as well as interface engineering. In this Perspective article, we examine the challenges and opportunities presented by spin textures in synthetic antiferromagnets and propose possible directions and prospects for future research in this burgeoning field.

* Author to whom correspondence should be addressed. Electronic mail: kang_wang@brown.edu (K. Wang), gang_xiao@brown.edu (G. Xiao).

## I.     INTRODUCTION

Recent advances in spintronics have led to the development of promising device prototypes that exploit the interplay between spin torques and spin textures. Spin textures such as magnetic domain walls and skyrmions hold great potential for encoding information in innovative electronic devices. The majority of research on spin textures has been centered on ferromagnetic films in which ferromagnetic layers interface with heavy elements possessing strong spin-orbit coupling.[1-4] Exchange interactions between neighboring spins of a ferromagnetic layer are mediated by a non-magnetic atom, resulting in the interfacial Dzyaloshinskii-Moriya interaction (DMI).[5,6] This interaction favors one sense of rotation of the magnetization over the other, thus manipulating the chirality of a spin texture. In addition, heavy elements can serve as spin Hall materials,[7-16] allowing for the conversion of charge current into spin current through spin-orbit coupling, thereby facilitating the efficient manipulation of spin textures. Despite recent progress, widespread implementation of such devices has been limited due to the inherent drawbacks of ferromagnetic materials. Their fringing magnetic fields restrict the density of closely-packed domain walls and impede the creation of ultrasmall skyrmions. The operation speed is restricted by their low-frequency precession. Topological spin textures like skyrmions experience a transverse deflection in motion[17,18] that complicates their efficient manipulation. Conversely, antiferromagnetic spin textures have the potential to overcome many of these limitations, leading to faster, smaller, more energy-efficient, and more robust electronics.

Historically, considerable research efforts in spintronics have been dedicated to ferromagnetic materials, while antiferromagnets were initially deemed "extremely interesting from a theoretical viewpoint" but with little practical value (see the Nobel lecture of Louis Néel who discovered antiferromagnetism in the 1930s).[19] This perspective remained unchanged for about half a century until the late 1980s when ferromagnetic/antiferromagnetic exchange bias[20,21] was utilized in spin valves, which were commercialized in hard disk recording heads. More recently, several properties of antiferromagnets have been discovered that take advantage of exchange enhancement, spin-orbit coupling, and the interplay between spin torques and spin textures. These properties render antiferromagnetic materials outstanding candidates for next-generation spintronic applications, leading to a new paradigm of antiferromagnetic spintronics.[22-28]

Synthetic antiferromagnets (SAFs) consist of two or more magnetic layers separated by spacers,[29] as illustrated in Fig. 1. These magnetic layers can be metallic ferromagnets,[30-44] diluted magnetic semiconductors,[45] or ferrimagnets.[46,47] The spacers, on the other hand, may consist of metallic,[30-35,37,38,40,43,44,46,47] semiconducting,[45] and insulating[36,39,41,42] layers. Interactions between spins in the magnetic layers and electrons in the spacer cause spin-dependent Friedel-like spatial oscillations in the spin density of the spacer. According to the Ruderman-Kittel-Kasuya-Yosida (RKKY) theory,[48-50] this leads to an interlayer ferromagnetic or antiferromagnetic coupling between magnetic layers. The RKKY-type interlayer exchange coupling (IEC) oscillates with the spacer thickness.[30-32,38,43,51,52] Table I provides a summary of the IEC strength and oscillation period for various spacers.[30,39-42,45] Additionally, the thickness of magnetic layers can also modulate quantum interferences inside magnetic layers, affecting the spin asymmetry of reflection coefficients at the interfaces and thereby quantum interferences in the spacer.[34-36,41,44,45,52-58]

SAFs, featuring an antiferromagnetic coupling between magnetic layers, resemble crystal antiferromagnets in that they do not produce stray fields, making them highly suitable for reducing device size and ensuring device stability against external fields. Additionally, SAFs exhibit high-frequency spin dynamics, making them attractive for high-speed electronics. Since the interlayer RKKY interaction is much weaker than the direct exchange or superexchange interaction in crystal antiferromagnets, SAFs can be manipulated more easily by engineering structures with different materials and interfaces. This enables the realization of diverse static and dynamic properties in SAFs, thereby enhancing their performance and capabilities across various applications, as summarized in Table II. [59-74]

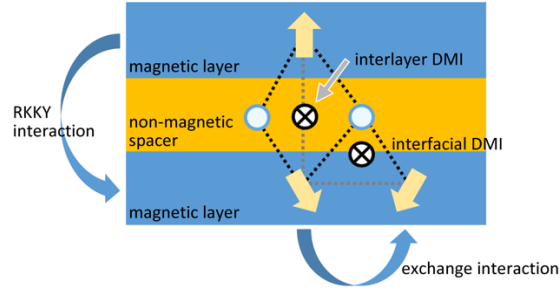

**FIG. 1.** A schematic representation of interaction energies in a SAF.

**TABLE I.** A summary of the IEC strength and oscillation period for various specific spacers. $A_1$ corresponds to the spacer thickness at the position of the first peak in antiferromagnetic IEC as the spacer thickness is increased. $J_1$ represents the IEC strength at this first peak, and $P$ denotes the oscillation period.

| Spacer | Properties | $A_1$ (Å) | $J_1$ (mJ/m²) | $P$ (Å) | Ref. |
|---|---|---|---|---|---|
| V | Metal | 9 | 0.1 | 9 | [30] |
| Cr | Metal | 7 | 0.24 | 18 | [30] |
| Cu | Metal | 8 | 0.3 | 10 | [30] |
| Nb | Metal | 9.5 | 0.02 | - | [30] |
| Mo | Metal | 5.2 | 0.12 | 11 | [30] |
| Ru | Metal | 3 | 5 | 11 | [30] |
| Rh | Metal | 7.9 | 1.6 | 9 | [30] |
| Ta | Metal | 7 | 0.01 | - | [30] |
| W | Metal | 5.5 | 0.03 | - | [30] |
| Re | Metal | 4.2 | 0.41 | 10 | [30] |
| Ir | Metal | 4 | 1.86 | 9 | [30] |
| Fe$_{37}$Cr$_{63}$ | Metal | - | 0.5 | - | [40] |
| GaAs:Be @5 K | Semiconductor | - | 0.00005 | - | [45] |
| MgO | Insulator | 25 | 0.004 | - | [41] |
| CaRu$_{0.5}$Ti$_{0.5}$O$_3$ | Insulator | 12 | 0.025 | - | [42] |
| VO$_2$ @200 K | Insulator | 7.5 | 0.05 | - | [39] |

**TABLE II.** Static and dynamic properties of SAFs in diverse applications.

| Applications | Static/Dynamic Properties | Ref. |
|---|---|---|
| Magnetic random access memory | Spin-transfer torque-driven magnetization reversal | [59,60] |
| | Spin-orbit torque-driven magnetization reversal | [61-63] |
| | Voltage-controlled magnetization reversal | [64] |
| Racetrack memory | Domain-wall motion | [65-68] |
| | Skyrmion motion | [66,69] |
| Spin-torque oscillators | Spin precession | [70] |
| | Gyrating motion of a skyrmion | [71] |
| Synapses | Voltage-controlled skyrmion-size variation | [72] |
| Neurons | Skyrmion motion through a barrier | [73] |
| True random number generator | Random motion of skyrmions through a Y-shaped branch | [74] |

Central to many of these applications is the intricate interplay between spin torques and magnetization. In addition to current-induced spin-transfer torque (STT)[59,60] and spin-orbit torque (SOT),[61-63] the application of an electric field can efficiently alter material properties by mediating carrier density at the Fermi surface.[38,75-78] Furthermore, magnon-mediated torque, which does not rely on conducting electrons, presents a promising avenue for the manipulation of magnetization.[79] This approach offers a potential solution to the energy-dissipation issue and enables more efficient control.

The observation of spin textures, such as domain walls[65,80,81] and skyrmions,[82-86] in SAFs has pointed to opportunities for further exploration and utilization. Although conceptual devices have been proposed, as illustrated in Table II,[65-69,71-74] several challenges remain unresolved, regarding the stabilization, efficient manipulation, and electrical detection of spin textures, as well as the development of three-dimensional spin textures. This perspective article aims to explore the challenges and opportunities associated with spin textures in SAFs. We present potential directions and prospects for future research in this field, aiming to identify strategies to harness the unique properties of SAFs to improve the performance of electronic devices that employ spin textures.

## II.    SPIN TEXTURES IN SYNTHETIC ANTIFERROMAGNETS

The formation of a spin texture is a result of the complex interplay between several energy terms, including anisotropy ($U_{ani}$),[87,88] exchange interaction ($U_{ex}$), interfacial DMI ($U_{IF-DMI}$),[5,6,89] Zeeman ($U_{Zeeman}$), and demagnetization ($U_{demag}$) energies. Additionally, the RKKY interaction ($U_{RKKY}$) in SAFs and the interlayer DMI ($U_{IL-DMI}$) in specific systems,[90-96] must also be considered. The free energy of a system can be expressed as

$$U = U_{ani} + U_{ex} + U_{IF-DMI} + U_{RKKY} + U_{IL-DMI} + U_{Zeeman} + U_{demag}. \qquad (1)$$

In this equation, $U_{ani} = -\sum_{\langle \alpha, \beta \rangle, \langle i, j \rangle} K_\alpha S_{\alpha,i}^{z}{}^2$ where $K$ is the anisotropy constant, the subscript $\alpha$ or $\beta$ represents the $\alpha$-th and $\beta$-th magnetic layers, and $i$ or $j$ represents the $i$-th and $j$-th spins of a magnetic layer, respectively. $U_{\text{ex}} = -\sum_{\langle \alpha, \beta \rangle, \langle i, j \rangle} J_{\alpha, ij} \mathbf{S}_{\alpha, i} \cdot \mathbf{S}_{\alpha, j}$ where $J_{\alpha, ij}$ is the exchange stiffness. $U_{\text{IF}-\text{DMI}} = -\sum_{\langle \alpha, \beta \rangle, \langle i, j \rangle} \mathbf{D}_{\alpha, ij} \cdot \left( \mathbf{S}_{\alpha, i} \times \mathbf{S}_{\alpha, j} \right)$ where $\mathbf{D}_{\alpha, ij}$ is the interfacial DMI vector. $U_{\text{RKKY}} = -\sum_{\langle \alpha, \beta \rangle, \langle i, j \rangle} J_{\alpha\beta, i} \mathbf{S}_{\alpha, i} \cdot \mathbf{S}_{\beta, i}$ where $J_{\alpha\beta, i}$ is the interlayer RKKY interaction between the $\alpha$-th and $\beta$-th magnetic layers. $U_{\text{IL}-\text{DMI}} = -\sum_{\langle \alpha, \beta \rangle, \langle i, j \rangle} \mathbf{D}_{\alpha\beta, i} \cdot \left( \mathbf{S}_{\alpha, i} \times \mathbf{S}_{\beta, i} \right)$ where $\mathbf{D}_{\alpha\beta, i}$ is the interlayer DMI between magnetic layers. $U_{\text{Zeeman}} = -\mu \sum_{\langle \alpha, \beta \rangle, \langle i, j \rangle} \mathbf{B}_{\text{ext}} \cdot \mathbf{S}_{\alpha, i}$ and $U_{\text{demag}} = -\mu \sum_{\langle \alpha, \beta \rangle, \langle i, j \rangle} \mathbf{B}_{\text{demag}} \cdot \mathbf{S}_{\alpha, i}$ where $\mathbf{B}_{\text{ext}}$ and $\mathbf{B}_{\text{demag}}$ are the external field and demagnetization field, respectively. Figure 1 provides a schematic representation of these interaction energies, which collectively determine the spin-texture configuration in SAFs.

Spin textures such as magnetic domain walls and skyrmions can emerge in a system through a spin-reorientation transition. To induce the emergence of spin textures in a ferromagnetic film, the magnetic anisotropy must be adjusted from in-plane to out-of-plane. However, in a SAF composed of two magnetic layers, it is not necessary to precisely adjust the magnetic anisotropy of each layer to achieve the transition from in-plane to perpendicular. Spin textures can emerge when the magnetic anisotropies of the two layers compensate each other, as illustrated in Figs. 2(a) – 2(c). This provides greater controls for creating spin textures in SAFs.

We use a macrospin model to compute the magnetization $M_z$ as a function of the perpendicular magnetic field $B_z$ by minimizing the free energy. In this model, we consider only two spins, each located in a different magnetic layer. Although the calculations do not provide information about spin textures, they provide insight into how a system undergoes the spin-reorientation transition. In our calculations, we adopt saturation magnetizations $M_{S1} = M_{S2} = 1.06 \text{ A m}^{-1}$, thicknesses of the two magnetic layers $d_1 = 0.8 \text{ nm}$, $d_2 = 1.0 \text{ nm}$, and the interlayer RKKY interaction strength $J_{12} = -1.07 \times 10^{-5} \text{ J m}^{-2}$, while neglecting the interlayer DMI. These magnetic parameters are based on our previous measurements of a CoFeB/Ta/CoFeB system.[44] Figure 2(b) shows the slope of $M_z$ with respect to $B_z$ at zero field, which indicates that the system undergoes a spin-reorientation transition when magnetic anisotropies of the two layers, $K_1$ and $K_2$, compensate each other.

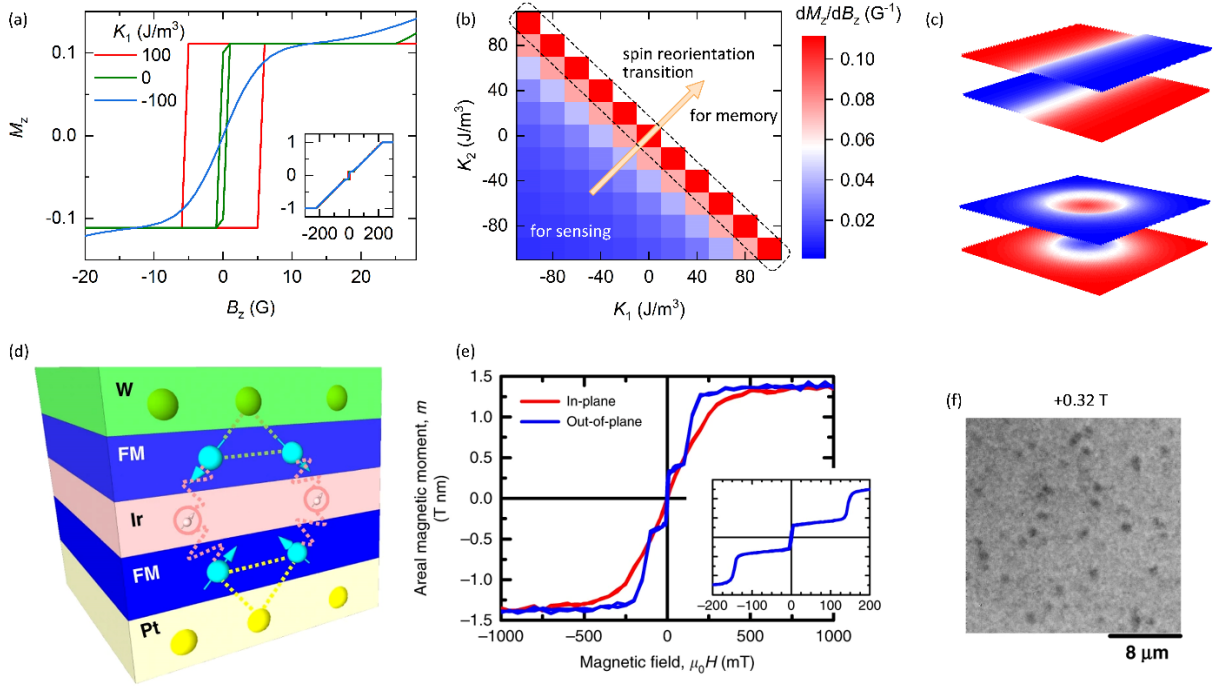

**FIG. 2.** (a) Numerical calculations of hysteresis loops for $M_z$ versus $B_z$ using different $K_2$ parameters while $K_1$ was kept at zero. (b) The slope of $M_z$ with respect to $B_z$ at zero field, which indicates that the system undergoes a spin-reorientation transition when magnetic anisotropies of the two layers, $K_1$ and $K_2$, compensate each other. (c) Schematic representations of domain walls and skyrmions in SAFs. (d) Schematic of stack structure of Pt/Co/Co$_{0.19}$Fe$_{0.56}$B$_{0.25}$/Ir/Co/CoFeB/W highlighting the interlayer antiferromagnetic coupling and interfacial DMI. (e) The magnetization behavior of the multilayer in responses to both in-plane (red) and out-of-plane (blue) magnetic fields. (f) A magneto-optic Kerr effect microscopy image of magnetic skyrmions at a magnetic field of 0.32 T. (d) to (e) are reproduced with permission from Dohi et al., Nat. Commun. **10**, 5153 (2019).[82] Copyright 2019 Springer Nature.

Domain walls and skyrmions can exist in a wide range of magnetic systems. These spin textures have been observed in multiple SAFs including Co/Ni/Co/Ru/Co/Ni/Co,[65,80] Co/Ni/Co/Rh/Co/Ni/Co,[81] Pt/Co/Ru/Co/W,[67] Pt/Co/Co$_{0.19}$Fe$_{0.56}$B$_{0.25}$/Ir/Co/CoFeB/W,[82] [Co/Pd]/Ru/[Co/Pd],[83] [Co/Pd]/Ir/Pt/[Co/Pd],[84] [Pt/Co/Ni/Co/Ru/Pt/Co/Ni/Co/Ru]$_2$,[85] and Pt/Co/Ru/Pt/Co/Ru,[86] as summarized in Table III. Figures 2(d) – 2(f) showcase the magnetization behavior of the Pt/Co/Co$_{0.19}$Fe$_{0.56}$B$_{0.25}$/Ir/Co/CoFeB/W multilayer in responses to in-plane and out-of-plane magnetic fields.[82] The results demonstrate a spin-reorientation transition within the system, and notably, the observation of magnetic skyrmions in the SAF. While magnetic anisotropy is a crucial parameter for the emergence of spin textures, the detailed configuration of a spin texture, such as the skyrmion size and domain-wall width, is strongly dependent on other material parameters as well. For example, Legrand et al. has shown a strong dependence of the skyrmion size on the interfacial DMI.[86] While some studies have been conducted on the origin and morphology of these spin textures, further research is necessary. A general theory on the skyrmion size and domain-wall width is needed, analogous to what has been developed for ferromagnetic systems.[97]

**TABLE III.** Summary of different SAF systems that host spin textures, along with their corresponding velocities $v_d$ at the current density $J_C$.

| Systems | Spin textures | Skyrmion size (nm) | $v_d$ (m/s) | $J_C$ (A/cm$^2$) | Ref. |
|---|---|---|---|---|---|
| Co/Ni/Co/Ru/Co/Ni/Co | domain walls | - | 750 | $3 \times 10^8$ | [65] |
| Co/Ni/Co/Ru/Co/Ni/Co | domain walls | - | 250 | $1.2 \times 10^8$ | [80] |
| Co/Ni/Co/Rh/Co/Ni/Co | domain walls | - | 550 | $2.2 \times 10^8$ | [81] |
| Co/CoFeB/Ir/Co/CoFeB | skyrmions | 1000 | 10 | $1.0 \times 10^8$ | [82] |
| [Co/Pd]/Ru/[Co/Pd] | skyrmions | 160 | - | - | [83] |
| [Co/Pd]/Ir/Pt/[Co/Pd] | skyrmions | 1000 | 0.5 | $2.3 \times 10^7$ | [84] |
| [Pt/Co/Ni/Co/Ru/Pt/Co/Ni/Co/Ru]$_2$ | skyrmions | $100-350$ | - | - | [85] |
| Pt/Co/Ru/Pt/Co/Ru | skyrmions | 50 | - | - | [86] |

Skyrmions, were only observed in a nonzero, finite field range in these SAFs. Zero-field stabilization of skyrmions is crucial for skyrmionic applications. Field-free creation and manipulation of ferromagnetic skyrmions have been verified by engineering interfacial exchange bias,[98,99] modifying the device geometry,[100,101] as well as by using external stimuli such as strain engineering,[102] chemisorption and desorption.[103] Zero-field creation and manipulation of skyrmions in SAFs remains elusive. Given that SAFs are immune to the external field, it is thereby difficult to create zero-field antiferromagnetic skyrmions through exchange-bias engineering. Instead, a biasing interaction provided by an additional magnetic layer may help to create zero-field skyrmions.[86] Moreover, external stimuli such as the strain, voltage, optics, and magnons, may also help to create zero-field skyrmions. The key is to switch magnetization in a defined region to form a magnetic bubble which may then be transformed into a skyrmion stabilized by the antisymmetric DMI.

Due to their inherent resistance to external fields and the absence of dipolar stray fields, spin textures in SAFs are widely recognized as superior to their ferromagnetic counterparts in terms of storage and stability of magnetic states. However, a significant challenge lies in the deterministic detection of spin textures in SAFs, which is critical for the successful implementation of electronic devices. Presently, various techniques are employed for the direct imaging of spin textures in SAFs, including the magneto-optic Kerr effect (MOKE) microscopy,[82,84] Lorentz transmission electron microscopy (L-TEM),[83] magnetic force microscopy (MFM),[86] and x-ray magnetic circular dichroism scanning transmission x-ray microscopy (XMCD-STXM).[85] Additionally, spin textures in SAFs can also be electrically detected through various methods such as the anomalous Hall effect[104] and tunneling magnetoresistance (TMR) effect.[105] Anomalous Hall detection requires using different thicknesses for the two magnetic layers to ensure that the magnetization cannot be fully compensated. Ma et al. demonstrated an anomalous Hall detection of skyrmion nucleation in arrays of magnetic nanodots built in a Hall cross, as depicted in Fig. 3.[104] Furthermore, Guang et al. have shown the electrical detection of skyrmions in ferromagnetic [Pt/Co/Ta] multilayers using the TMR effect in a magnetic tunnel junction.[105] However, the deterministic detection of a single skyrmion in SAFs using these methods remains a challenge for ongoing research. The key lies in creating a single skyrmion in a controlled manner and reducing electronic noise. Single skyrmions can be generated by passing current through a constricted geometry,[1,100,106,107] exploiting local defects,[100,108,109] or applying localized spin torque,[3,110-112] as demonstrated in ferromagnetic systems[1,3,100,106-111] as well as in micromagnetic simulations[112] of SAFs. To reduce skyrmion noise, the use of local pinning centers can be implemented, which can also reliably position skyrmions and avoid skyrmion annihilation. Additionally, the electronic signal can be enhanced by reducing the detection area relative to the skyrmion size.

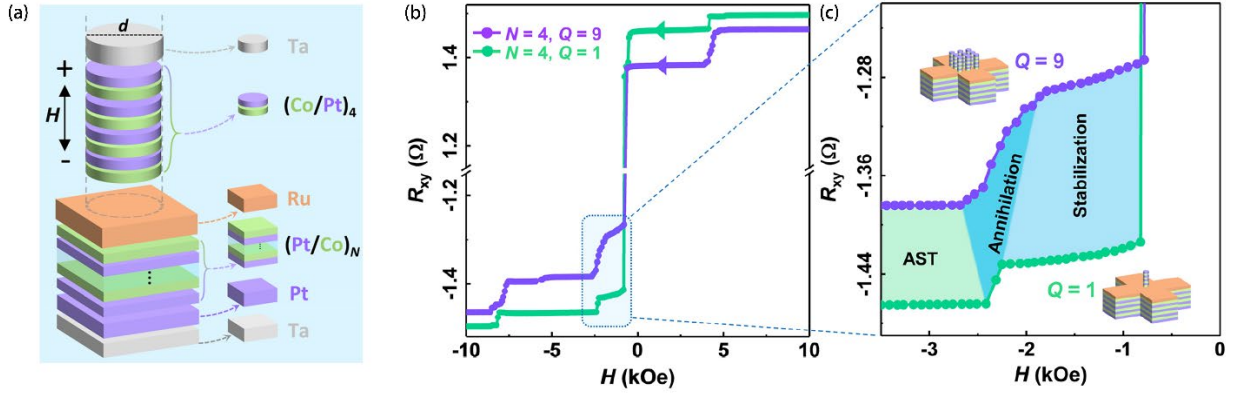

**FIG. 3.** (a) Schematic illustration of nanostructured SAF multilayers. (b) Descending branches of the anomalous Hall resistance $R_{xy}$ as a function of the magnetic field for SAF multilayers with $N = 4$, $Q = 1$ and $Q = 9$. Here, $N$ represents the number of repeats of the bottom layer in the multilayer structure, while $Q$ corresponds to the number of nanodots present in a Hall cross configuration. (c) Zoomed-in view of the specific region highlighted by the blue frame in (b). In this region, the presence of "antiferromagnetic spin textures" is denoted by the abbreviation "AST". Reproduced with permission from Ma et al. Appl. Phys. Rev. 9, 021404 (2022).[104] Copyright 2022 AIP Publishing.

## III.    DYNAMICS OF SPIN TEXTURES

Spin textures in SAFs are particularly intriguing given their high-speed motion which makes them effective carriers for spintronic devices. The dynamics of a spin is governed by the Landau-Lifshitz-Gilbert (LLG) equation,

$$\dot{\mathbf{m}} = -\frac{\gamma_L}{M_S} \mathbf{m} \times \left(-\frac{\delta U}{\delta \mathbf{m}}\right) + \alpha_L \mathbf{m} \times \dot{\mathbf{m}} + \tau. \tag{2}$$

Here, $\gamma_L$ and $\alpha_L$ represent the gyromagnetic ratio and Gilbert damping constant, respectively. $\tau$ refers to additional torques, such as the conventional STT $\tau_{FL,STT} = \frac{p a^3 \gamma_L \hbar}{2 e M_S} (\mathbf{j} \cdot \nabla \mathbf{m})$, $\tau_{DL,STT} = -\frac{p a^3 \gamma_L \hbar}{2 e M_S} \beta [\mathbf{m} \times (\mathbf{j} \cdot \nabla \mathbf{m})]$, and current-induced SOTs $\tau_{FL,SOT} = \frac{\gamma_L \Theta_{SH} \hbar p}{2 e M_S d} (\mathbf{m}_p \times \mathbf{m})$, $\tau_{DL,SOT} = \frac{\gamma_L \Theta_{SH} \hbar p}{2 e M_S d} \alpha_L [\mathbf{m} \times (\mathbf{m}_p \times \mathbf{m})]$. The STT can be understood through a simple model – when the current passes through a magnetic system, the electron spin is deflected by the magnetic moment, and at the same time there is a reactive force from the electrons to the magnetic moment, which is the STT as schematically illustrated in

Fig. 3. In STT expressions, $e$ is the elementary charge, $\hbar$ is the reduced Planck constant, $a$ is the lattice constant, $p$ is the spin polarization, $\mathbf{j}$ is the charge current, $\beta$ is the adiabatic constant. The SOT arises from spin current generated by the spin-orbit coupling effect such as the Rashba effect[113] and spin Hall effects.[7,10,13,14,16,114-118] The spin-orbit coupling causes electrons with different spin polarizations to deflect in different directions, resulting in a spin current, as schematically shown in Fig. 4. Considering the polarization direction $\mathbf{m}_\text{p}$ of the spin current and the charge current $\mathbf{j}$, the spin current is presented as $\mathbf{j}_\text{S} = \frac{\hbar}{2e}\Theta_\text{SH}(\mathbf{m}_p \times \mathbf{j})$ where $\Theta_\text{SH}$ is the spin Hall angle characterizing the conversion efficiency. A spin texture refers to a collection of spins, which can be effectively manipulated as an individual, particle-like entity.

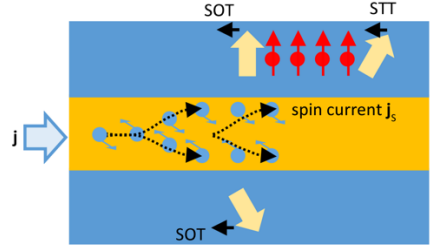

**FIG. 4.** Schematic representations of STT and SOTs in a multilayer.

In a ferromagnetic system, the presence of a threshold driving force causes a spin texture, such as the domain wall, to precess. This results in an abrupt decrease in domain-wall velocity known as the Walker breakdown.[119,120] However, this phenomenon is absent in antiferromagnetic systems due to the symmetry of the torques acting on domain walls in the two sublattices.[121] This effect, together with an additional effective torque arising from the RKKY interaction $-\frac{\gamma_\text{L}}{M_\text{S}}\mathbf{m} \times \left(-\frac{\delta}{\delta\mathbf{m}}U_\text{RKKY}\right)$,[65,122] leads to a higher effective mobility of domain walls in SAFs, as presented in Figs. 5(a) and 5(b). For similar reasons, one would also expect other spin textures, such as skyrmions, to exhibit high speeds of motion in SAFs, as illustrated in Fig. 5(c).

When considering a rigid spin texture, the Thiele equation can be derived by integrating the LLG equation in space to describe the motion of a spin texture under a driving force.[123,124] In the case of a conventional STT, the Thiele equation is expressed as

$$\boldsymbol{\mathcal{G}} \times (\mathbf{v}_{s1} - \mathbf{v}_{d1}) + \overleftrightarrow{\boldsymbol{\mathcal{D}}}(\beta \mathbf{v}_{s1} - \alpha_{\mathrm{L}} \mathbf{v}_{d1}) + \mathbf{F}_1 + \mathbf{F}_{2/1} = 0 \tag{3a}$$

$$-\boldsymbol{\mathcal{G}} \times (\mathbf{v}_{s2} - \mathbf{v}_{d2}) + \overleftrightarrow{\boldsymbol{\mathcal{D}}}(\beta \mathbf{v}_{s2} - \alpha_{\mathrm{L}} \mathbf{v}_{d2}) + \mathbf{F}_2 + \mathbf{F}_{1/2} = 0. \tag{3b}$$

In the case of current-induced SOTs, the Thiele equation can then be written as

$$\boldsymbol{\mathcal{G}} \times \mathbf{v}_{d1} + \alpha_{\mathrm{L}} \overleftrightarrow{\boldsymbol{\mathcal{D}}} \mathbf{v}_{d1} - \frac{\gamma_{\mathrm{L}} \Theta_{\mathrm{SH}} \hbar p}{2 M_s e d} \boldsymbol{\mathcal{T}} + \mathbf{F}_1 + \mathbf{F}_{2/1} = 0 \tag{4a}$$

$$-\boldsymbol{\mathcal{G}} \times \mathbf{v}_{d2} + \alpha_{\mathrm{L}} \overleftrightarrow{\boldsymbol{\mathcal{D}}} \mathbf{v}_{d2} + \frac{\gamma_{\mathrm{L}} \Theta_{\mathrm{SH}} \hbar p}{2 M_s e d} \boldsymbol{\mathcal{T}} + \mathbf{F}_2 + \mathbf{F}_{1/2} = 0. \tag{4b}$$

$\boldsymbol{\mathcal{G}} = (0, 0, \mathcal{G})$ is the gyrovector and $\mathcal{G} = \int \mathbf{m} \cdot \left( \frac{\partial \mathbf{m}}{\partial x} \times \frac{\partial \mathbf{m}}{\partial y} \right) dx dy$ is related to the topological property of a spin texture. $\overleftrightarrow{\boldsymbol{\mathcal{D}}}$ is the dissipative tensor with $\mathcal{D}_{ij} = \frac{Md}{\gamma_{\mathrm{L}} a} \int \left( \frac{\partial \mathbf{m}}{\partial \xi_i} \cdot \frac{\partial \mathbf{m}}{\partial \xi_i} \right) dx dy$. $\mathbf{F} = -\frac{\gamma_{\mathrm{L}}}{M_s d} \boldsymbol{\nabla} U$ is the pinning force arising from a non-uniform distribution of material parameters. In Eq. 3, $\mathbf{v}_s = -\frac{p a^3 \gamma_{\mathrm{L}} \hbar}{2 e M_s} \mathbf{j}$ is the effective spin velocity, and $\mathbf{v}_d$ is the velocity of a spin texture. In Eq. 4, $\boldsymbol{\mathcal{T}} = \int \mathrm{d}^2 x j \mathbf{m}_p \cdot (\boldsymbol{\nabla} \mathbf{m} \times \mathbf{m})$ represents the driving force from the spin current. $\mathbf{F}_{2/1}$ and $\mathbf{F}_{1/2}$ represent the coupling forces that describe the interaction between the textures in two magnetic layers.

Trivial spin textures, such as domain walls, have a topological charge of zero. Therefore, the first terms in Eqs. 3 and 4 vanish, and these textures do not experience transverse motion. On the other hand, non-trivial spin textures, like skyrmions, have a nonzero topological charge and experience a Magnus force (the first terms in Eqs. 3 and 4) when in motion. This results in the skyrmion Hall effect, which describes the additional transverse motion that accompanies the longitudinal motion induced by the current.[17,18] In applications such as racetrack memory devices, the intrinsic skyrmion Hall effect may drive a texture toward a racetrack boundary, resulting in its annihilation. Interestingly, in SAFs, the skyrmion Hall effect may disappear because of the opposing Magnus forces acting on textures in the two magnetic layers, as displayed in Fig. 5(d).[82,84]

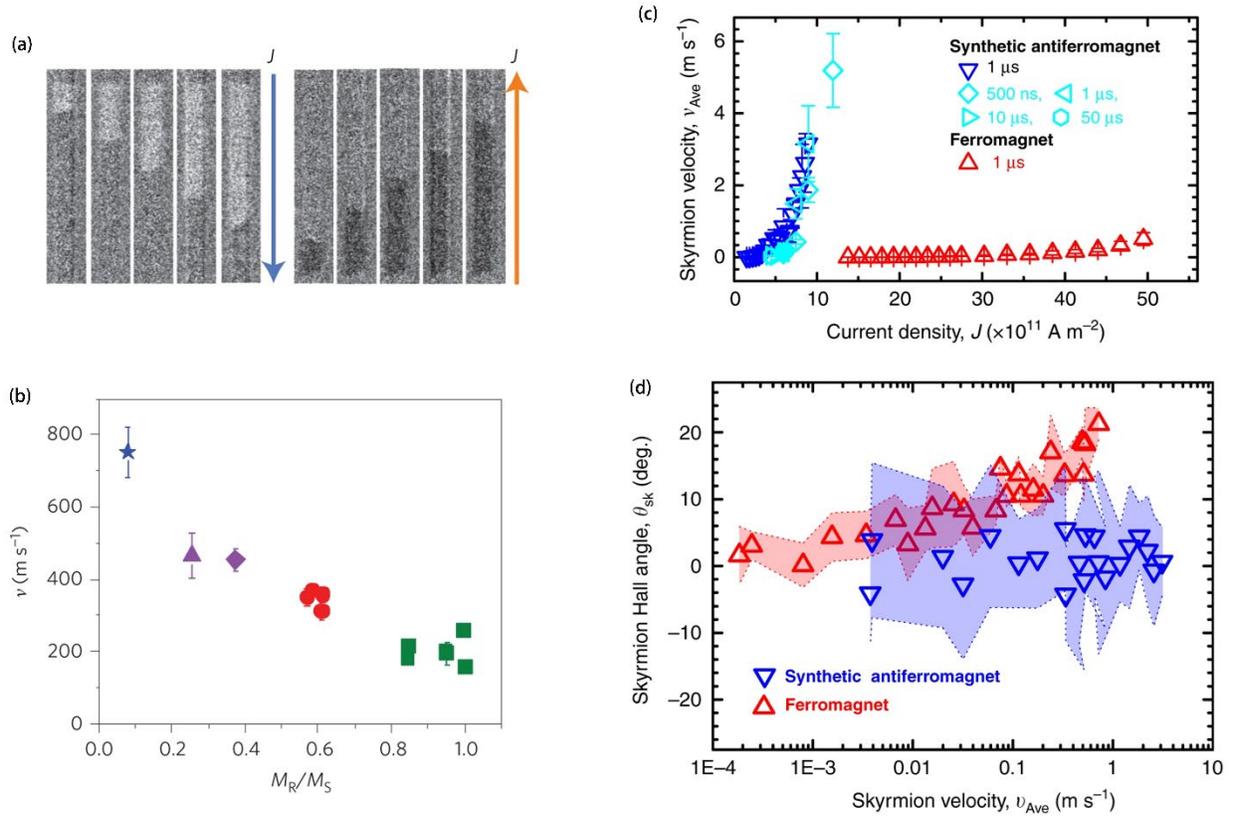

**FIG. 5.** (a) The magneto-optic Kerr effect microscopy images illustrating the movement of a single domain-wall along a SAF nanowire composed of Co/Ni/Co/Ru/Ni/Co. (b) The domain-wall velocity dependent on the ratio of remnant magnetization $M_R$ to saturation magnetization $M_S$. (c) The average velocity of skyrmion bubbles in a SAF Pt/Co/Co$_{0.19}$Fe$_{0.56}$B$_{0.25}$/Ir/Co/CoFeB/W multilayer (blue) and a ferromagnetic Pt/Co/Co$_{0.19}$Fe$_{0.56}$B$_{0.25}$/Ir/Ru multilayer (red). (d) The skyrmion Hall angle as a function of the average skyrmion velocity for skyrmion bubbles in both SAF and ferromagnetic layers. (a) and (b) are reproduced with permission from Yang et al. Nat. Nanotechnol. 10, 221 (2015).[65] Copyright 2015 Springer Nature. (c) and (d) are reproduced with permission from Dohi et al. Nat. Commun. 10, 5153 (2019).[82] Copyright 2019 Springer Nature.

The advancement of spintronic technology depends heavily on the effective manipulation of spin textures. Based on the analysis presented above, several approaches could be considered to further improve the efficiency of electrical manipulation.[125,126] The speed of a skyrmion can be increased by using the non-adiabatic STT $\beta$ and confining it in a one-dimensional channel.[126] In this scenario, the velocity of the skyrmion motion is proportional to $\beta/\alpha_L \mathbf{v}_s$. Therefore, one way to improve the efficiency of electrical manipulation is to enhance the non-adiabatic STT $\beta/\alpha_L$ which is inversely proportional to the product of the spin splitting $2\varepsilon$ and the partial density of states of magnetic atoms at the Fermi level $N_M(\varepsilon_F)$. An unexpectedly large $|\beta/\alpha_L|$ value of approximately 8.3 is obtained in FeGe. This remarkable finding implies that the injection of current can efficiently propel skyrmions in FeGe at speeds of approximately 8.3 m/s per MA/cm$^2$, surpassing the conventional limitations associated with the low mobility in current-driven skyrmion motion. Furthermore, this achievement reaches the mobility value of 10 m/s per MA/cm$^2$ as proposed by A. Fert et al.,[110] highlighting FeGe's potential as a material for developing low-power and high-speed spintronic technologies.[125-128] It is of great importance to identify other materials with a large $\beta/\alpha_L$. This effect has also been noted for a skyrmion bundle with zero topological charge, which moves along the current direction without the need for an artificial one-dimensional channel.[125] For similar reasons, materials with a high $\beta/\alpha_L$ value are expected to be beneficial for enhancing the speed of textures in SAFs, which warrants further investigation.

Compared to SOTs, STT generally has a weaker effect on manipulating textures in films. Another approach to improve the efficiency of electrical control is to enhance the charge-to-spin conversion of the spin Hall materials adjacent to magnetic layers, thus improving the SOT generation efficiency. The spin Hall effect can be caused by either the intrinsic mechanism of a non-zero Berry curvature or the impurity-induced skew scattering or side jump. Heavy elements with strong spin-orbit coupling such as Pt,[7-9] $\beta$-Ta,[10,11] and $\beta$-W[12,13] have demonstrated large spin Hall effects. Additionally, investigating topological materials with spin-orbit coupling and spin-momentum locking in surface and/or bulk states can improve the intrinsic Berry curvature-induced spin Hall effect. In addition to topological insulators such as Bi$_2$Se$_3$,[129] Cr-doped (Bi$_{0.5}$Sb$_{0.5}$)$_2$Te$_3$,[130] Bi$_{1.5}$Sb$_{0.5}$Te$_{1.7}$Se$_{1.3}$,[131] and Weyl/Dirac semimetals WTe$_2$[132-135] that have been verified in experiments, other materials including Weyl/Dirac semimetals TaAs,[136] IrO$_2$,[137] W$_3$Ta,[138] PtTe$_2$,[139] ZrSiTe,[140] LaAlSi and LaAlGe[141] have been theorized to be intriguing potential candidates. Furthermore, alloying can be utilized to enhance extrinsic scattering.[16,142-151] This may, however, also lead to an increase in the longitudinal resistivity and thus may not necessarily improve the efficiency of electrical manipulation. Therefore, an active and ongoing area of research is to identify materials with high SOT generation efficiency while maintaining high conductivity. Alternatively, researchers are exploring new effects such as the orbital Hall effect,[152] which is prominent in $3d$ elements with low resistivity, as well as the intrinsic SOT generation in magnetic materials.[153-158]

The spin current, typically associated with charge flow, may generate non-negligible Joule heating over time along with a short nanometer-scale propagation length.[28] A magnon current, which describes the precessing motion of spin moments, successfully addresses this shortcoming with a longer propagation length of several micrometers. Not only must we first understand the principles of this recently discovered form of spin current, but we must also study how the magnon current interacts with spin textures. Han et al. has experimentally shown magnon-driven domain-wall motion in ferromagnetic Pt/[Co/Ni]$_9$/Ru multilayers.[159] Conversely, spin textures can serve as knobs to tune spin waves. Wagner et al. demonstrated the channeling of spin waves inside nano-sized domain walls in a 40-nm-thick Ni$_{81}$Fe$_{19}$ film.[160] In addition, Hämäläinen et al. has controlled spin-wave propagation across domain walls by modifying the domain-wall configuration.[161] While 90° head-to-head or tail-to-tail domain walls are transparent to spin waves, head-to-tail domain-wall configuration hinders spin-wave propagation. Moreover, theoretical works have predicted that domain walls can also be utilized to manipulate the phase of spin waves.[162-166] More recently, Han et al. has experimentally verified such a phase-shifting effect in ferromagnetic Pt/[Co/Ni]$_9$/Ru multilayers.[159] How exactly spin waves interact with spin textures in SAFs remains unexplored, although this may be studied in ways similar to their ferromagnetic counterparts. Research on this topic may provide a new avenue to efficiently manipulating spin textures and to controlling spin waves in SAFs, leading to advancements in non-volatile memory and magnonic logic devices. This effect is anticipated to be particularly significant in insulator-based structures.

Moreover, the utilization of alternative stimuli, such as bias voltage, also presents a promising avenue for achieving energy-efficient control over spin textures in SAFs. Recent research has shown that applying an electric field[38,75] can efficiently modify the material properties of SAFs by influencing the carrier density at the Fermi surface.[38,77,78] Notably, Kossal et al. successfully demonstrated a continuous transition from antiferromagnetic to ferromagnetic coupling, and vice versa, using a gate voltage,[76] which has implications for the manipulation of spin textures in SAFs. Additionally, Guan et al. have presented a noteworthy manipulation of current-induced domain-wall motion in SAFs through ionic liquid gating, as depicted in Fig. 6.[80] However, achieving energy-efficient manipulation of spin textures in SAFs still requires further investigation and is the topic of ongoing research. The application of bias voltage has shown promise in inducing skyrmion creation, motion, and deletion in ferromagnetic films by manipulating material parameters.[167-171] This suggests the feasibility of applying similar concepts to the context of SAFs, thereby opening up new possibilities for their control and utilization.

To achieve the efficient manipulation of spin textures, it is important to not only improve the spin torque generation efficiency, but also to fabricate films with higher uniformity of material parameters. This can help reduce the pinning force $\mathbf{F} = -\frac{\gamma_L}{M_S d} \nabla U$ that impedes the motion of spin textures which ultimately leads to a lower velocity of skyrmion motion in SAFs reported thus far. It should be noted, however, that the pinning effect can also be useful for improving the thermal stability of spin textures. Achieving a balance between the effective manipulation and thermal stability remains a challenge. One potential solution is to use artificial pinning centers to guide motion and prevent the annihilation of spin textures in SAFs.[68] Artificial pinning centers such as voids in multilayers, thickness modulations, embedded impurity atoms and adatoms adhering to the surface may be formed with advanced thin film fabrication, lithography, irradiation, ion implantation, and laser ablation techniques.

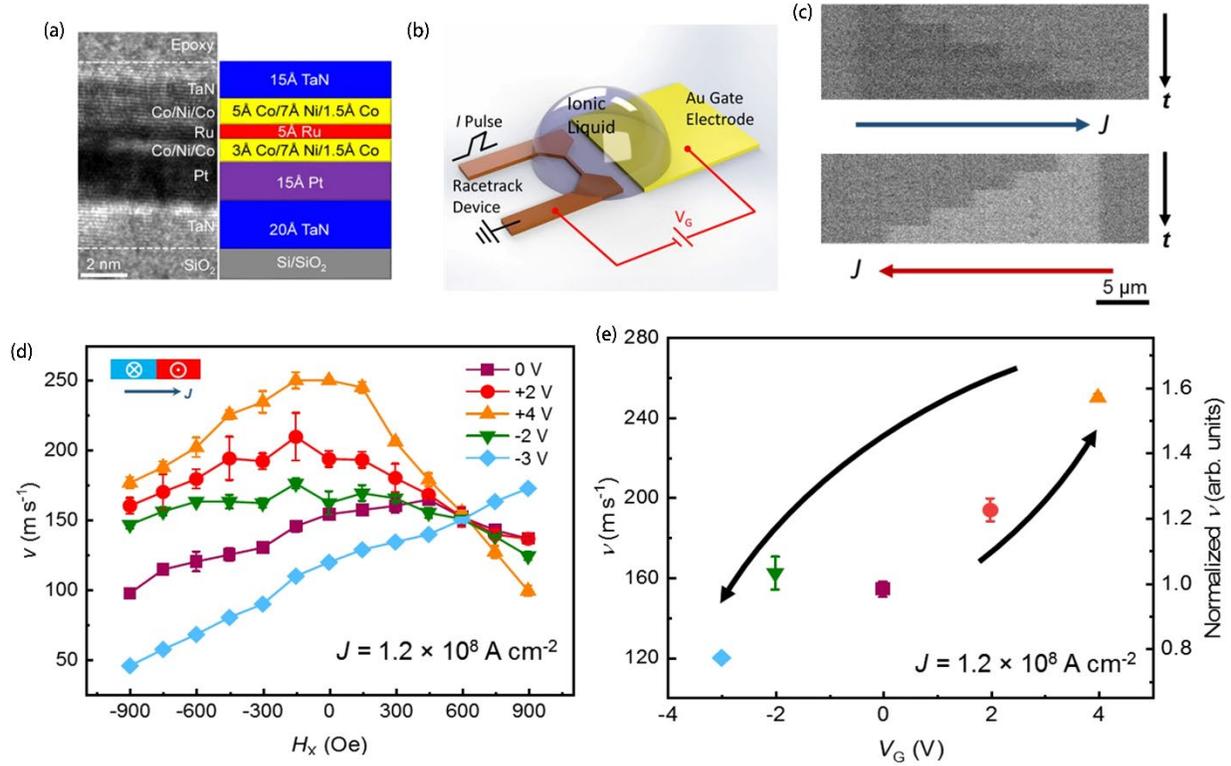

**FIG. 6.** (a) Cross-section transmission electron microscopy (TEM) image (left) and schematic structure (right) for a SAF composed of Co/Ni/Co/Ru/Co/Ni/Co. (b) Sketch of racetrack device with gate electrode. (c) The magneto-optic Kerr effect (MOKE) microscopy images of a single domain wall moving along the nanowire after applying sequences of $4 \times 10$ ns pulses with current density of $J = 1.2 \times 10^8$ A cm$^{-2}$. (d) Dependence of domain-wall velocity on the in-plane magnetic field $H_x$ for various gate voltages $V_G$. The in-plane magnetic field is applied along the racetrack. (e) The domain-wall velocity as a function of $V_G$ at $H_x = 0$. Reproduced with permission from Guan et al. Nat. Commun. 12, 5002 (2021).[80] Copyright 2021 Springer Nature.

## IV.    TOWARDS THREE-DIMENSIONAL SPIN TEXTURES

Currently, most of the research efforts on spin textures focus on two-dimensional magnetic films. Introducing a third dimension may lead to the discovery of new physical phenomena and functionalities. For example, three-dimensional spin textures could be used for ultra-high-density data storage and may exhibit nontrivial dynamics which are key to developing next-generation spintronic devices.

There are two ways to introduce a third dimension to magnetic films. One approach is to introduce curvature into the film. Curved magnetic films have significant potential for applications, especially in high-density data storage and processing. By using curvature as a new design parameter, it is possible to create new interactions that influence the morphology and dynamics of the spin textures.[172-175]

An alternative method is to build extended systems with intricate magnetic interactions in the third dimension, as schematically shown in Fig. 7. The formation of kink defects in the third dimension can be achieved by creating superlattices with varying IEC and magnetic layer thicknesses.[176,177] These defects can be introduced and propagated by external field pulses. Recently, a theoretical work has predicted the existence of an interlayer DMI in a RKKY-coupled magnetic film with an in-plane asymmetry.[90] The interlayer DMI governs a chiral magnetization across magnetic multilayers, which has been observed in some prototype systems.[91-95] The interlayer DMI originates either from crystalline asymmetry[90] or from the in-plane asymmetry of the IEC.[92,94,95] The interlayer and the intralayer DMIs provide two effective parameters to design and construct three-dimensional magnetic structures and devices. A notable example of a three-dimensional spin texture is the topologically protected hopfion, which places high demands on the system in which it exists. Kent et al. have shown that hopfions can be created in nanoscale disks.[178] The presence of interlayer DMI provides a new pathway for generating hopfions in extended systems which can be more easily manipulated.

Currently, the most promising techniques for detecting and directly observing three-dimensional spin textures appear to be x-ray microscopy/tomography[179-182] and off-axis electron holography/tomography.[178,183] These methods involve capturing two-dimensional transmission images from different angles, which are then utilized to reconstruct the three-dimensional structure of the targeted object. Recent advancements have showcased the effectiveness of these three-dimensional tomographic imaging techniques in studying various spin textures, including magnetic singularities[180] and the skyrmion string.[179] This demonstrates the applicability of such techniques in studying and analyzing complex three-dimensional spin textures with high precision and accuracy.

The electrical detection of three-dimensional spin textures presents an ongoing challenge in current research. Surface-sensitive techniques, like the TMR effect, face limitations in detecting spin textures located within the bulk of multilayers. However, alternative methods such as anomalous Hall or topological Hall detections show promise in electrically detecting three-dimensional spin textures. Particularly, the locally uncompensated emergent field of a hopfion can generate a topological Hall signal,[184] although this phenomenon still requires experimental examination to confirm its existence and feasibility. Addressing these challenges and exploring novel electrical detection techniques will be crucial for advancing our understanding and utilization of three-dimensional spin textures.

Constructing three-dimensional spin textures presents a challenge that requires multilayer fabrication with interlayer DMI which significantly impacts magnetization. Fabricating multilayers with a large antisymmetric/symmetric exchange interaction ratio can improve the contribution from the crystalline asymmetry. Wedge-shaped samples can improve the in-plane asymmetry of the IEC.

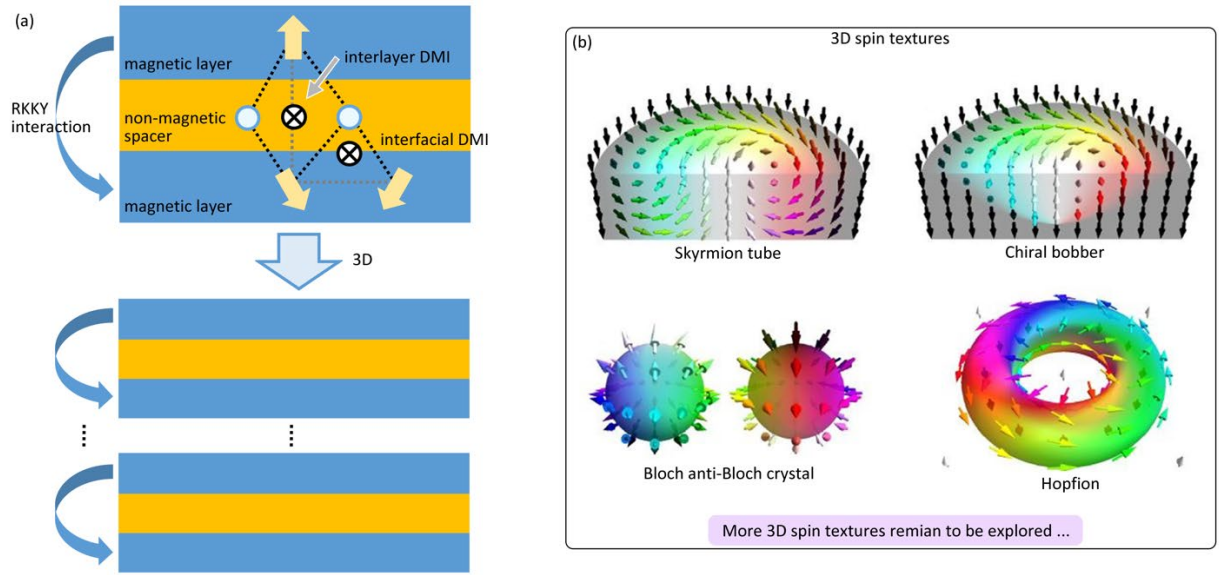

**FIG. 7.** (a) A schematic representation of extended multilayers with intricate magnetic interactions in the third dimension. (b) Schematic representations of three-dimensional spin textures.[185] Reproduced with permission from Göbel et al., Phys. Rep. 895, 1-28 (2021). Copyright 2021 Elsevier.

Chirality-induced nonreciprocity, in which a system responds differently to opposite excitations, is a universal phenomenon. Fernández-Pacheco et al.[91] and Han et al.[92] have shown an asymmetric response of the interlayer DMI-governed chiral magnetization to the external magnetic field. Moreover,

Wang et al.[94] and Masuda et al.[95] have observed an asymmetric current-driven switching of the chiral magnetization which they attribute to the spin torque from the interlayer DMI. While the interactions of single-domain structures with external stimuli have been extensively studied, the response of spin textures to stimuli is not well understood and requires further investigation. These spin structures may show nontrivial responses to external stimuli which will undoubtedly expand the scope of spintronic applications.

## V.     CONCLUSIONS

In this perspective article, we highlight the advantages of spin textures in SAFs over their ferromagnetic counterparts including smaller size, higher mobility, and vanishing skyrmion Hall effect. We also discuss the potential of using magnetic multilayers, which combine both interfacial and interlayer DMIs as a promising platform for constructing three-dimensional spin textures. We examine how to incorporate a significant interlayer DMI into SAFs, which can enhance their nontrivial responses to external stimuli. Spin textures have the potential to revolutionize next-generation beyond-CMOS data storage,[103,186,187] logic,[188-190] probabilistic computing,[191-194] and neuromorphic computing[195] devices. For a deeper understanding, we recommend referring to focused reviews on specific topics.[196,197] The ongoing developments of spin textures in SAFs are expected to greatly expand the scope of spintronic applications.

## ACKNOWLEDGEMENTS

G.X. acknowledges funding support from the National Science Foundation (NSF) under Grant No. DMR-2202514.